\documentclass[12pt]{iopart}
\usepackage{epsfig}
\usepackage{citesort}

\newcommand{\gtrsim}{\,\rlap{\lower3.7pt\hbox{$\mathchar\sim$}}
\raise1pt\hbox{$>$}\,}
\newcommand{\lesssim}{\,\rlap{\lower3.7pt\hbox{$\mathchar\sim$}}
\raise1pt\hbox{$<$}\,}

\newcommand{\hMpc}{$h\,\rm Mpc^{-1}$}

\begin{document}

\title{Cosmological mass limits on neutrinos, axions,
and other light particles}
\author{Steen Hannestad}
\address{Department of Physics, University of Southern Denmark\\
Campusvej 55, DK-5230 Odense M, Denmark}
\author{Georg Raffelt}
\address{Max-Planck-Institut f\"ur Physik
(Werner-Heisenberg-Institut)\\
F\"ohringer Ring 6, 80805 M\"unchen, Germany}

\date{{\today}}

\begin{abstract}
  The small-scale power spectrum of the cosmological matter
  distribution, together with other cosmological data, provides a
  sensitive measure of the hot dark matter fraction, leading to
  restrictive neutrino mass limits. We extend this argument to generic
  cases of low-mass thermal relics. We vary the cosmic epoch of
  thermal decoupling, the radiation content of the universe, and the
  new particle's spin degrees of freedom. Our treatment covers various
  scenarios of active plus sterile neutrinos or axion-like particles.
  For three degenerate massive neutrinos, we reproduce the well-known
  limit of $m_\nu<0.34$~eV.  In a 3+1 scenario of 3~massless and
  1~fully thermalized sterile neutrino we find $m_\nu<1.0$~eV.
  Thermally produced QCD axions must obey $m_a<3.0$~eV, superseding
  limits from a direct telescope search, but leaving room for solar
  eV-mass axions to be discovered by the CAST experiment.
\end{abstract}
\maketitle

\section{Introduction} 

The observed universe is surprisingly well described by a simple
Friedmann-Robertson-Walker model that is characterized by a handful of
well-measured global parameters. Together with the gravitational
instability theory for the formation of structure, a ``concordance
model'' has emerged that does not seem to conflict with any firmly
established observations. Of course, the physical nature of most of
the gravitating matter and energy remains elusive, except that most of
the dark matter must be of the cold variety to avoid excessive free
streaming in the early universe that would suppress small-scale power
of the observed matter distribution. In the framework of this standard
paradigm, the observed power spectrum of the cosmic matter
distribution provides restrictive limits on the hot dark matter
fraction of the cosmological matter~cocktail~\cite{Hu:1997mj}.

Taking advantage of the latest cosmological data, notably the 2dF and
SDSS galaxy redshift surveys and the WMAP measurement of the cosmic
microwave background (CMB) temperature fluctuations, several authors
have used this argument to derive restrictive neutrino mass limits of
$\sum m_\nu<0.7$--1.0~eV (95\% CL)
\cite{Elgaroy:2002bi,Hannestad:2002xv,Hannestad:2003xv,%
  Elgaroy:2003yh,Spergel:2003cb,Allen:2003pt,Barger:2003vs}.  The
exact value of the nominal limit depends on the included data sets and
on the assumed priors for some of the cosmological parameters.
However, the small spread between different nominal limits suggests
that the universe is a robust laboratory for constraining the overall
neutrino mass scale, and perhaps eventually for positively determining
its value \cite{Hannestad:2002cn,Abazajian:2002ck,Kaplinghat:2003bh,%
  Bashinsky:2003tk}.  Of course, these limits or a future positive
detection depend on the validity of the standard cosmological
paradigm. For example, the spectrum of primordial density fluctuations
may not follow a simple power law, modifying all of these results.
Still, very recently it was shown that the neutrino mass limit is
surprisingly robust against a contribution of nonadiabatic, incoherent
fluctuations such as would be predicted by topological
defects~\cite{Brandenberger:2004kc}.

We are here concerned with a different extension of the standard
picture in that we study the effect of a non-standard number density
of neutrinos, perhaps caused by the existence of additional neutrinos
such as sterile states, or new low-mass particles such as axions.  One
of us has previously shown that increasing the assumed number density
of neutrinos actually {\em weakens} the mass limit, contrary to naive
intuition, because the increased radiation density counteracts the
effect of the increased hot dark matter
fraction~\cite{Hannestad:2003xv}.  Therefore, the usual neutrino mass
limits do not trivially translate into more general cases where the
hot dark matter fraction and the radiation content are varied
independently.

The free-streaming effect of low-mass particles on the power
spectrum of the matter distribution and of the CMB temperature
fluctuations depends on the particle mass $m_X$, their number
density and velocity distribution, and on the radiation content in
other degrees of freedom that do not carry mass.  We will consider
two generic classes of cases. One covers neutrinos, i.e.\
fermionic degrees of freedom with a momentum distribution given by
that of ordinary neutrinos. We vary the mass $m_\nu$, the
effective number of flavors carrying that mass, and the effective
number of massless flavors. Our second generic case is that of
particles that were once in thermal equilibrium (thermal relics)
so that their contribution to the hot dark matter fraction and
their velocity distribution is perfectly defined by the cosmic
temperature $T_{\rm D}$ when they thermally decoupled from the
background medium, by their mass, and by their number of internal
degrees of freedom.

In Sec.~2 we describe our general methodology for deriving the allowed
range of non-standard cosmological parameters.  In Sec.~3 we apply our
method to various scenarios involving active and sterile neutrinos.
In Sec.~4 we consider general thermal relics, and finally we conclude
in Sec.~5.

\section{Likelihood Analysis}  

\subsection{Theoretical Predictions}

In order to derive limits on the parameters characterizing the new
particles we compare theoretical power spectra for the matter
distribution and the CMB temperature fluctuations with observational
data in analogy to the previous works by one of
us~\cite{Hannestad:2002xv,Hannestad:2003xv}.  The predicted power
spectra are calculated with the publicly available CMBFAST
package~\cite{CMBFAST}. This package allows one to include directly
the effective number of massless neutrino degrees of freedom and of
neutrinos carrying a mass as well as the cosmic hot dark matter
fraction, assumed to be equally distributed among the massive flavors.
Put another way, the standard version of the code assumes a velocity
distribution of the neutrinos that ordinary active neutrinos with a
mass would have. For our more general case of arbitrary thermal relics
we modified the code accordingly.

As a set of cosmological parameters apart from the new particle
characteristics we choose the matter density $\Omega_m$, the baryon
density $\Omega_b$, the Hubble parameter $H_0$, the scalar spectral
index of the primordial fluctuation spectrum $n_s$, the optical depth
to reionization $\tau$, the normalization of the CMB power spectrum
$Q$, and the bias parameter~$b$.  We restrict our analysis to
geometrically flat models $\Omega = \Omega_m + \Omega_\Lambda = 1$.
Some of the other parameters are not kept entirely free, but are
varied over a prior range that is determined from cosmological
observations other than CMB and LSS.  In flat models the matter
density is restricted by observations of Type Ia supernovae to be
$\Omega_m = 0.28 \pm 0.14$ \cite{Perlmutter:1998np}.  Furthermore, the
HST Hubble key project has measured $H_0=72 \pm 8~{\rm km}~{\rm
  s}^{-1}~{\rm Mpc}^{-1}$ \cite{Freedman:2000cf}.  The cosmological
parameters and their assumed priors are listed in
Table~\ref{tab:priors}.  The actual marginalization over parameters
was performed using a simulated annealing
procedure~\cite{Hannestad:wx}. The cosmological parameters in our
model corresponds to the simplest $\Lambda$CDM model which fits
present data. While our results would be changed slightly by including
additional parameters, there would be no qualitative changes. The
neutrino mass is mainly degenerate with the matter density, the bias
parameter, and to a lesser extent with the Hubble parameter, but not
for instance with the curvature~\cite{Hannestad:2002xv}. However, our
prior on $\Omega_m$ from SN~Ia data is not very restrictive, and the
results would not change significantly by removing it. Removing the
prior on $H_0$ also will not lead to any substantial change in our
results.

\begin{table}[ht]
\caption{\label{tab:priors} Priors on cosmological parameters used in
the likelihood analysis.}
\begin{indented}
\item[]
\begin{tabular}{@{}lll}
\br
Parameter &Prior&Distribution\cr
\mr
$\Omega=\Omega_m+\Omega_\Lambda$&1&Fixed\\
$\Omega_m$ & $0.28 \pm 0.14$&Gaussian\\
$h$ & $0.72 \pm 0.08$&Gaussian\\
$\Omega_b h^2$ & 0.014--0.040&Top hat\\
$n_s$ & 0.6--1.4& Top hat\\
$\tau$ & 0--1 &Top hat\\
$Q$ & --- &Free\\
$b$ & --- &Free\\
\br
\end{tabular}
\end{indented}
\end{table}

\subsection{Cosmological Data}

\subsubsection{Large Scale Structure (LSS).}

At present there are two large galaxy surveys of comparable size, the
Sloan Digital Sky Survey (SDSS) \cite{Tegmark:2003uf,Tegmark:2003ud}
and the 2dFGRS (2~degree Field Galaxy Redshift Survey) \cite{2dFGRS}.
Once the SDSS is completed in 2005 it will be significantly larger and
more accurate than the 2dFGRS. At present the two surveys are,
however, comparable in precision and in the present paper we use data
from the 2dFGRS alone.

Tegmark, Hamilton and Xu \cite{THX} have calculated a power spectrum,
$P(k)$, from this data, which we use in the present work. The 2dFGRS
data extends to very small scales where there are large effects of
non-linearity. Since we calculate only linear power spectra, we follow
standard procedures and use only data on scales larger than $k =
0.2\,$\hMpc, where effects of non-linearity should be
minimal (see for instance Ref.~\cite{Tegmark:2003ud} for a
discussion). With this cut the number of data points for the power
spectrum reduces to~18.

\subsubsection{Cosmic Microwave Background.}

The temperature fluctuations are conveniently described in terms of
the spherical harmonics power spectrum $C_{T,l} \equiv \langle
|a_{lm}|^2 \rangle$, where $\frac{\Delta T}{T} (\theta,\phi) =
\sum_{lm} a_{lm}Y_{lm}(\theta,\phi)$.  Since Thomson scattering
polarizes light, there are also power spectra coming from the
polarization. The polarization can be divided into a curl-free $(E)$
and a curl $(B)$ component, yielding four independent power spectra:
$C_{T,l}$, $C_{E,l}$, $C_{B,l}$, and the $T$-$E$ cross-correlation
$C_{TE,l}$.

The WMAP experiment has reported data only on $C_{T,l}$ and $C_{TE,l}$
as described in
Refs.~\cite{Spergel:2003cb,Bennett:2003bz,Kogut:2003et,%
Hinshaw:2003ex,Verde:2003ey,Peiris:2003ff}.  We have performed our
likelihood analysis using the prescription given by the WMAP
collaboration~\cite{Spergel:2003cb,Bennett:2003bz,Kogut:2003et,%
Hinshaw:2003ex,Verde:2003ey,Peiris:2003ff} which includes the
correlation between different $C_l$'s. Foreground contamination has
already been subtracted from their published data.

In addition we use other CMB data from the compilation by Wang {\it et
al.} \cite{wang3} which includes data at higher $l$.  Altogether this
data set has 28~points.

\newpage

\section{Neutrinos} 

\subsection{Mass Equally Distributed Among $N_\nu$ Species}

Our first class of cases consists of different scenarios with $N_\nu$
flavors of neutrinos with equal momentum distributions at those early
epochs where they are relativistic. Several of these species are
assumed to carry equal masses $m_\nu$.

The first and simplest example is that of $N_\nu$ flavors carrying
equal masses $m_\nu$.  In this case the hot dark matter fraction is
given by the standard expression
\begin{equation}
\Omega_\nu h^2=N_\nu\,\frac{m_\nu}{91.5~{\rm eV}}\,,
\label{eq:degenerate}
\end{equation}
where $h$ is the Hubble constant in units of $100~\rm
km~s^{-1}~Mpc^{-1}$.  Of course, the limiting case $N_\nu=3$
corresponds to ordinary neutrinos with a degenerate pattern of masses.

A value for $N_\nu$ much larger than 3 contradicts the observed
primordial helium abundance in the framework of the standard big-bang
nucleosynthesis (BBN) theory.  However, if the primordial $\nu_e$
chemical potential takes on a suitable value, its effect on the $n/p$
freeze-out abundance can always compensate for the expansion-rate
effect caused by the additional flavors.  Such a fine-tuned $\nu_e$
degeneracy parameter may not seem particularly plausible, but it
serves as one possibility to escape the BBN limits.  In any event, we
think it is useful to study what can be learned about neutrino
parameters from large-scale structure data alone, without combining
this information with data pertaining to an entirely different
cosmological epoch.

Since we know that the number of ordinary neutrino flavors is exactly
3, the additional degrees of freedom would have to be sterile states,
perhaps Dirac partners of the ordinary neutrinos that were thermally
equilibrated by some unknown mechanism.

Without sterile states, in principle the effective $N_\nu$ can be
enhanced by suitable neutrino chemical potentials for the ordinary
flavors.  However, the observed neutrino mixing parameters imply that
all three standard flavors reach full thermal equilibrium (i.e.\
kinetic and chemical equilibrium) before the BBN epoch so that the
restrictive BBN limit on the $\nu_e$ degeneracy parameter applies to
all flavors~\cite{Dolgov:2002ab}. Therefore, interpreting $N_\nu>3$ in
terms of an increased density of the ordinary flavors requires some
unspecified way to escape the BBN limits. We also note that our
treatment assumes a neutrino velocity distribution corresponding to a
non-degenerate thermal distribution. Therefore, a case with large
neutrino chemical potentials is not strictly covered here.

\begin{figure}[ht]
\hspace*{1.5cm}\includegraphics[width=70mm]{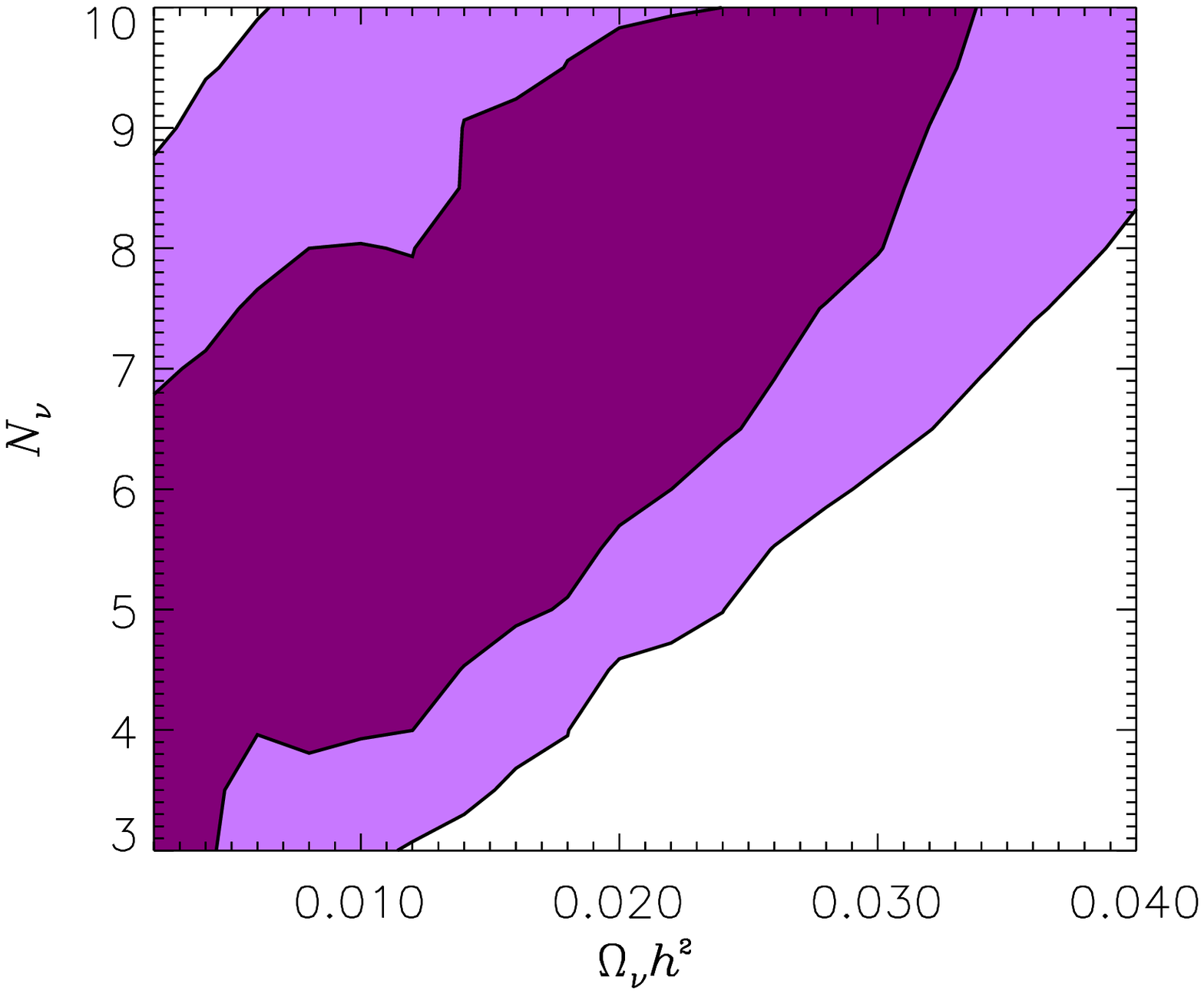}
\includegraphics[width=70mm]{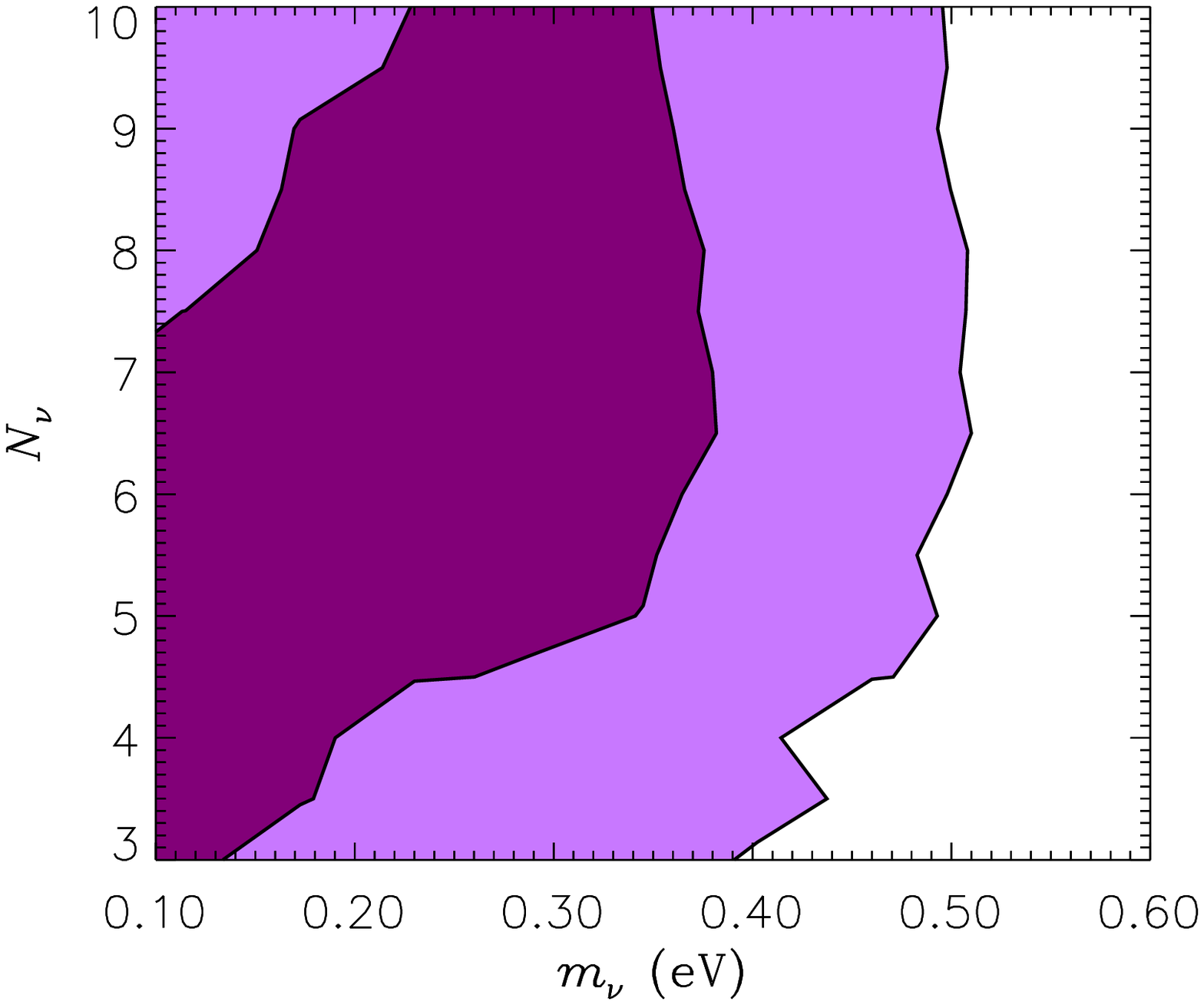}
\caption{Likelihood contours (68\% and 95\%) for the case of
$N_\nu$ neutrinos with equal masses $m_\nu$. Left panel:
\hbox{$N_\nu$-$\Omega_\nu h^2$-plane}. Right panel: Equivalent
\hbox{$N_\nu$-$m_\nu$-plane}.} \label{fig:neutrino1}
\end{figure}

After marginalizing over the cosmological parameters shown in
Table~\ref{tab:priors}, the 68\% and 95\% CL allowed region of
neutrino parameters is shown in Fig.~\ref{fig:neutrino1}. In the left
panel this allowed region is shown in the \hbox{$N_\nu$-$\Omega_\nu
h^2$-plane}, in the right panel the equivalent region in the
\hbox{$N_\nu$-$m_\nu$-plane}.

From the right panel of Fig.~\ref{fig:neutrino1} it is apparent that
the 95\% CL upper limit on $m_\nu$ is nearly independent of $N_\nu$,
contrary to naive intuition. One of us first observed this effect in
Ref.~\cite{Hannestad:2003xv}. The limits stated there agree with our
present ones that are given explicitly in Table~\ref{tab:masslimits1}
for several values of~$N_\nu$.

\begin{table}[ht]
\caption{\label{tab:masslimits1} Neutrino mass limits (95\% CL) for
$N_\nu$ flavors with equal mass $m_\nu$.}
\begin{indented}
\item[]
\begin{tabular}{@{}lll}
\br
Flavors&\multicolumn{2}{l}{Mass limits [eV]}\\
$N_\nu$&$N_\nu m_\nu$&$m_\nu$\\
\mr
3&1.01&0.34\\
4&1.38&0.35\\
5&2.12&0.42\\
6&2.69&0.45\\
\br
\end{tabular}
\end{indented}
\end{table}

If we specifically assume that the neutrinos are Dirac particles
with a degenerate mass spectrum, and that the right-handed
partners were thermally equilibrated by some unknown mechanism so
that they have equal number densities with the left-handed states,
the cosmological mass limit on $m_\nu$ degrades from 0.34~eV to
0.45~eV.

\subsection{Mass in One Fully Thermalized Species}

\begin{figure}
\hspace*{1.5cm}\includegraphics[width=70mm]{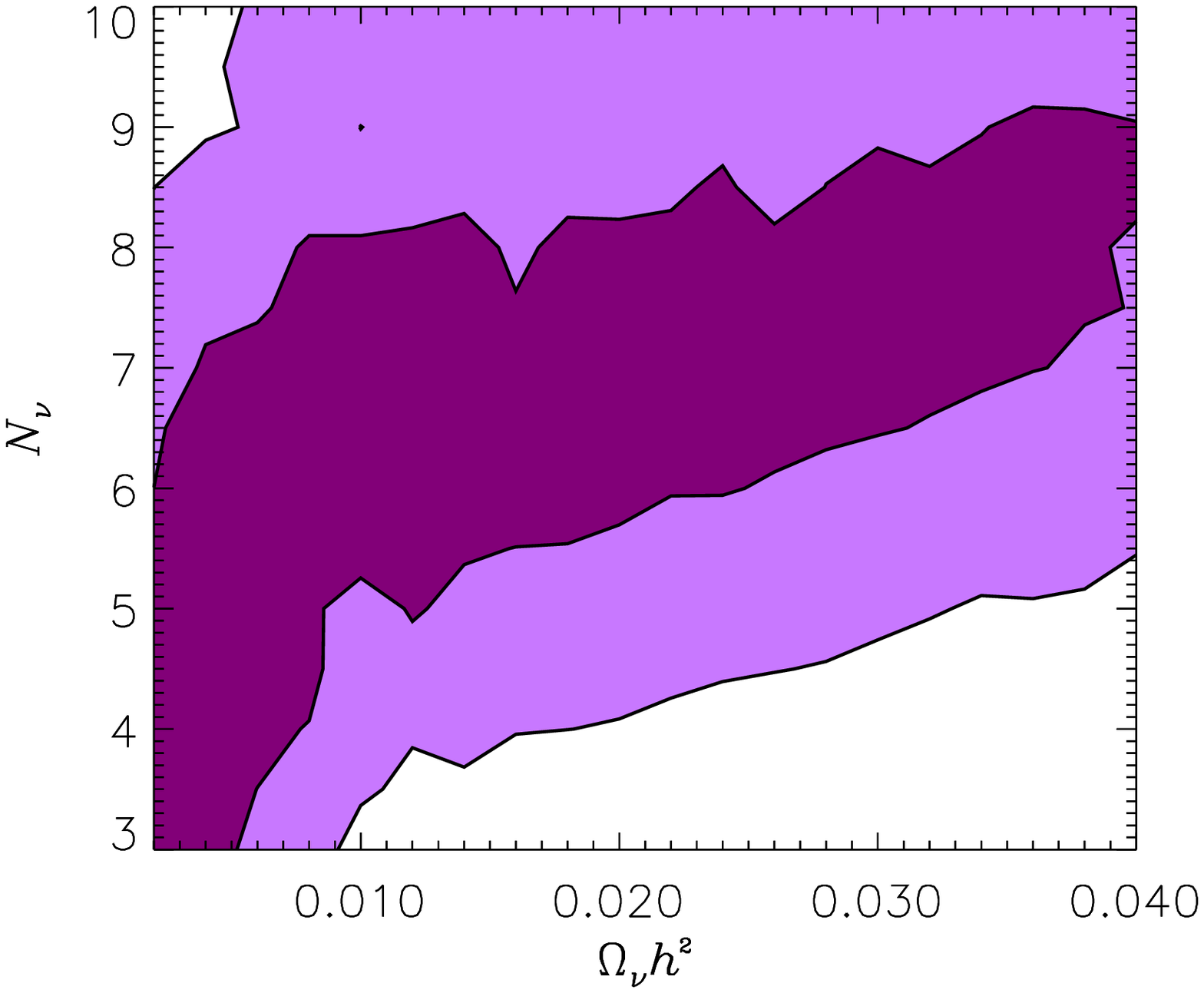}
\includegraphics[width=70mm]{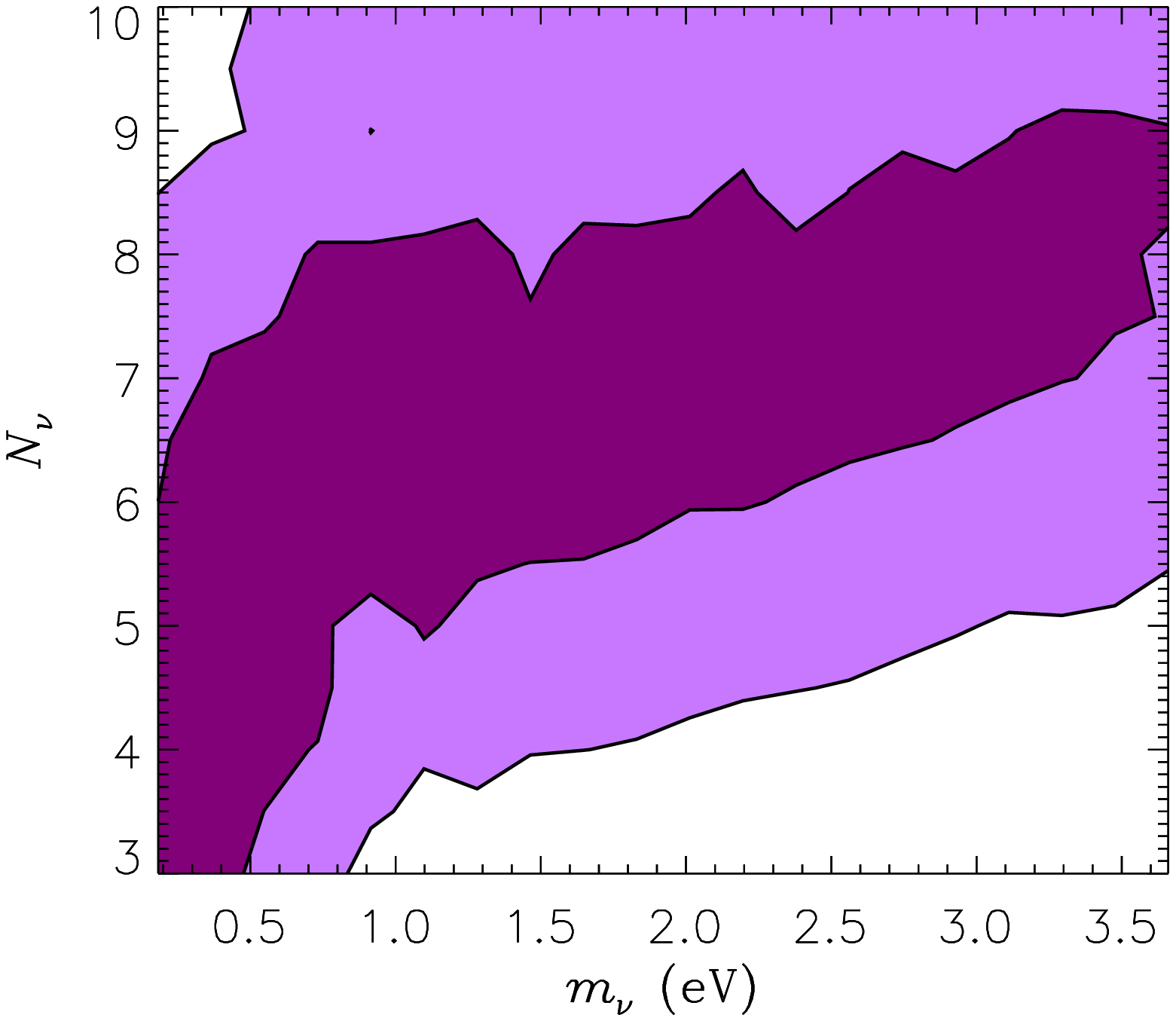}
\caption{Likelihood contours (68\% and 95\%)
  for the case of $N_\nu$ flavors with exactly one of them carrying a
  mass $m_\nu$. Left panel: \hbox{$N_\nu$-$\Omega_\nu h^2$-plane}.
  Right panel: Equivalent \hbox{$N_\nu$-$m_\nu$-plane}.
\label{fig:neutrino2}}
\end{figure}

As a next case we consider one fully excited neutrino flavor that
carries a mass $m_\nu$ so that the cosmic hot dark matter fraction is
given by
\begin{equation}
\Omega_\nu h^2=\frac{m_\nu}{91.5~{\rm eV}}\,.
\end{equation}
In addition there are $N_\nu-1$ massless species that contribute to
the radiation density.  The small mass differences between the
ordinary neutrinos imply that this scenario cannot be realized among
the ordinary neutrinos so that we need additional sterile states.  For
example, the 3+1 model for explaining the current neutrino oscillation
data including LSND would be a case in point.  As in the previous
case, one must assume that the BBN limits are circumvented by some
mechanism, for example a suitable $\nu_e$ chemical potential. Since we
have 3 flavors of ordinary neutrinos that would contribute to the
tally of massless states, and because one fully thermalized species is
assumed to carry a mass, the present scenario requires $N_\nu\geq4$.
The allowed region of $N_\nu$-$m_\nu$-values is shown in
Fig.~\ref{fig:neutrino2} where we have extended the plot down to
$N_\nu=3$.  In Table~\ref{tab:masslimits2} we give explicit numerical
95\% CL mass limits for several values of~$N_\nu$.

\begin{table}
\caption{\label{tab:masslimits2} Neutrino mass limits (95\% CL) for
$N_\nu$ flavors with exactly one of them carrying mass $m_\nu$.}
\begin{indented}
\item[]
\begin{tabular}{@{}ll}
\br
Flavors&Mass limit [eV]\\
\mr
3&0.73\\
4&1.05\\
5&2.47\\
6&4.13\\
\br
\end{tabular}
\end{indented}
\end{table}

\subsection{Three ``Massless'' Standard Neutrinos}

Our last neutrino example is one where we consider the three standard
neutrinos to have very small hierarchical masses, i.e.\ to be
effectively massless for our purpose. In addition we assume there to
be $N_\nu-3$ other flavors, e.g.\ sterile states, that carry equal
masses $m_\nu$ so that the hot dark matter fraction is
\begin{equation}
\Omega_\nu h^2=(N_\nu-3) \frac{m_\nu}{91.5~{\rm eV}}\,.
\end{equation}
An example would be a scenario with an additional sterile neutrino
that did not reach full thermal equilibrium so that it is present
with less than one full effective degree of freedom.

\begin{figure}[ht]
\hspace*{1.5cm}\includegraphics[width=70mm]{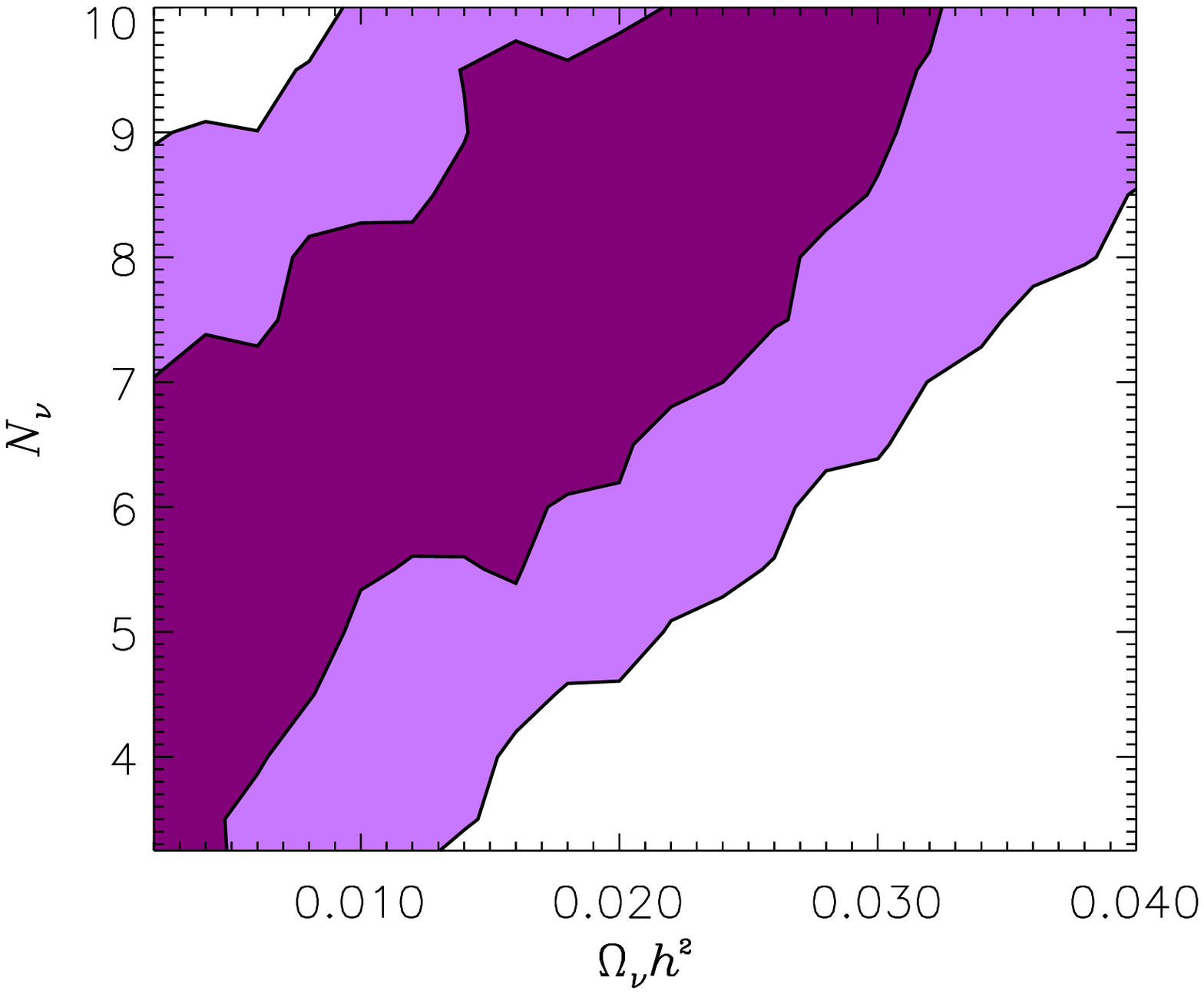}
\includegraphics[width=70mm]{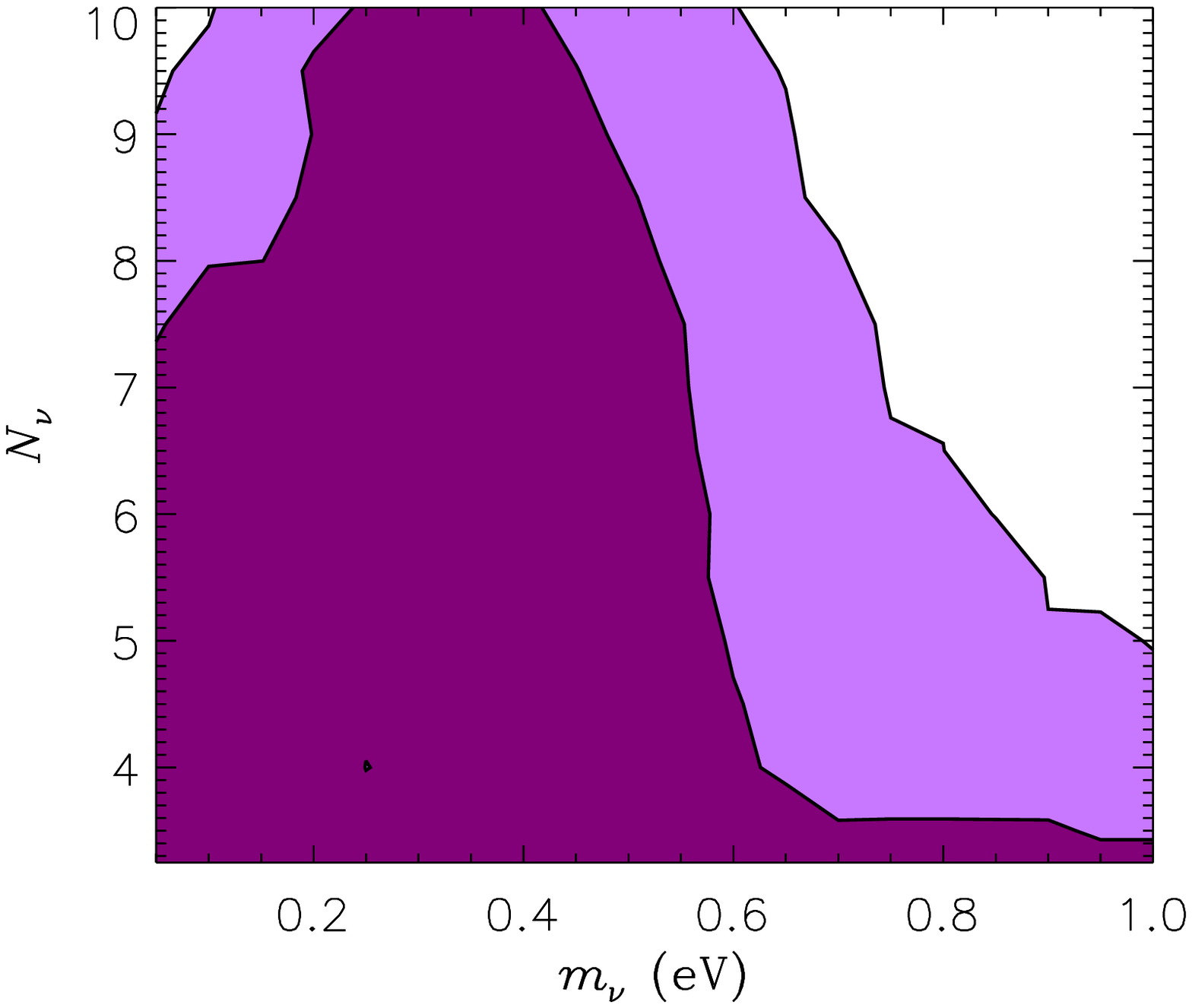}
\caption{\label{fig:neutrino3} Likelihood contours (68\% and 95\%)
for the case of $N_\nu$ flavors, $N_\nu-3$ each carrying mass
$m_\nu$, and an additional 3 massless flavors.}
\end{figure}

\begin{table}[ht]
\caption{Neutrino mass limits for the case of $N_\nu$ species,
with 3 massless flavors and $N_\nu-3$ massive neutrinos, each
carrying mass $m_\nu$.} \label{tab:masslimits3}
\begin{indented}
\item[]
\begin{tabular}{@{}ll}
\br
Flavors&Mass limit [eV]\\
\mr
3.25&2.95\\
3.50&1.67\\
3.75&1.21\\
4.00&1.02\\
4.50&0.87\\
5.00&0.82\\
5.50&0.77\\
6.00&0.71\\
\br
\end{tabular}
\end{indented}
\end{table}

The allowed region of $N_\nu$-$m_\nu$-values is shown in
Fig.~\ref{fig:neutrino3}. In Table~\ref{tab:masslimits3} we give
explicit numerical 95\% CL mass limits for several values
of~$N_\nu$. Note that the case $N_\nu=4$ is identical to the
$N_\nu=4$ case in the preceding section. The slight difference in
quoted mass limit (1.02 eV instead of 1.05 eV) is due to the
fact that the limit is interpolated from the calculated likelihood
grid.

Note that for less than one massive species the mass limit becomes
quite poor because the limit on $\Omega_\nu h^2$ remains almost
constant down to $N_\nu=3.25$, whereas the limit on $m_\nu$ is
proportional to $1/(N_\nu-3)$.

\subsection{Understanding the $\Omega_\nu$-$N_\nu$-degeneracy}

\begin{figure}[ht]
\hspace*{2.5cm}\includegraphics[width=90mm]{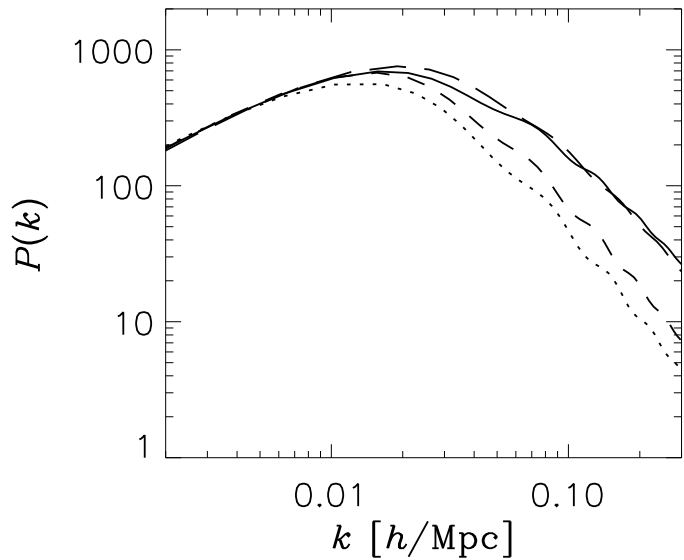}
\caption{\label{fig:pspec} Power spectra for $\Lambda$CDM
  models with $\Omega_b = 0.05$, $\Omega = 1$, $h=0.7$, $n_s=1$, and
  $N_{\nu,{\rm massive}}=1$ and a common large-scale normalization.
  The full line is for $\Omega_\nu=0$, $\Omega_m=0.25$, $N_\nu=3$,
  dashed is for $\Omega_\nu=0.05$, $\Omega_m=0.25$, $N_\nu=3$, dotted
  is for $\Omega_\nu=0.05$, $\Omega_m=0.25$, $N_\nu=8$, and
  long-dashed is for $\Omega_\nu=0.05$, $\Omega_m=0.35$, $N_\nu=8$.}
\end{figure}

In all of our cases there is a conspicuous degeneracy between
$\Omega_\nu$ and $N_\nu$. This can be understood by studying the power
spectra for different models. Figure~\ref{fig:pspec} shows power
spectra for various $\Lambda$CDM models with $\Omega_b = 0.05$,
$\Omega = 1$, $h=0.7$, $n_s=1$, and $N_{\nu,{\rm massive}}=1$, but
with different values of $\Omega_m$, $\Omega_\nu$, $N_\nu$, and $b$.
The full line corresponds to the standard $\Lambda$CDM model with no
neutrino mass. The dashed line is for $\Omega_\nu=0.05$,
$\Omega_m=0.25$, and $N_\nu=3$, and the power suppression below the
free-streaming scale is evident. The dotted line is for
$\Omega_\nu=0.05$, $\Omega_m=0.25$, and $N_\nu=8$. Here, the spectrum
is suppressed even further, at scales up to the horizon size at matter
radiation equality (because of later matter-radiation equality).
Finally the long-dashed line is for $\Omega_\nu=0.05$,
$\Omega_m=0.35$, and $N_\nu=8$. Increasing the matter density leads to
earlier matter domination, which in turn leads to less small-scale
suppression and smaller horizon size at matter radiation equality. The
power spectrum of this model is almost indistinguishable from the
standard $\Lambda$CDM model ($\Omega_\nu=0$, $N_\nu=3$), explaining
the $\Omega_\nu$-$N_\nu$-degeneracy.

\section{General Thermal Relics}

\subsection{Low-Mass Fermions}

Our second generic class of models is that of some new particle with
mass $m_X$ that was once in thermal equilibrium, but decoupled when
the universe had a temperature $T_{\rm D}$ and the radiation density
was characterized by $g_{*\rm D}$ effective thermal degrees of
freedom. The relevant masses are of order eV so that the particles
decouple when they are relativistic, i.e.\ at decoupling they are
characterized by a Fermi-Dirac or Bose-Einstein distribution of
temperature $T_{\rm D}$. The $N_\nu=3$ ordinary neutrinos are now
always assumed to be~massless.

In the present-day universe, the new particles will be
non-relativistic and contribute a matter fraction
\begin{equation}\label{eq:omegax}
\Omega_X h^2=\frac{m_X g_X}{183~{\rm eV}}\,
\frac{g_{*\nu}}{g_{*\rm D}}\times
\cases{1&fermions\cr{\textstyle\frac{4}{3}}&bosons\cr}
\end{equation}
where $g_X$ is the number of the particle's internal degrees of
freedom while $g_{*\nu}$ is the effective number of thermal degrees of
freedom when ordinary neutrinos freeze out with $g_{*\nu}=10.75$ in
the absence of new particles.  For the case of $N_\nu$ ordinary
neutrinos with equal masses (degenerate mass scenario) we have $g_X=2
N_\nu$ (two spin degrees for each flavors), $T_{\rm D}\approx 2~\rm
MeV$ and thus $g_{*\rm D}=g_{*\nu}$ so that Eq.~(\ref{eq:omegax})
corresponds to the usual result $\Omega_\nu h^2=N_\nu m_\nu/91.5~\rm
eV$ as stated already in Eq.~(\ref{eq:degenerate}).

In the late epochs that are important for structure formation, the
momentum distribution of the new particles is characterized by a
thermal distribution with temperature $T_X$ that is given by
\begin{equation}
\frac{T_X}{T_\nu}=\left(\frac{g_{*\nu}}{g_{*\rm D}}\right)^{1/3}
\,.
\end{equation}
We have modified CMBFAST to incorporate a fermionic thermal relic with
such a momentum distribution.

\begin{figure}
\begin{center}
\hspace*{1.5cm}
\includegraphics[width=70mm]{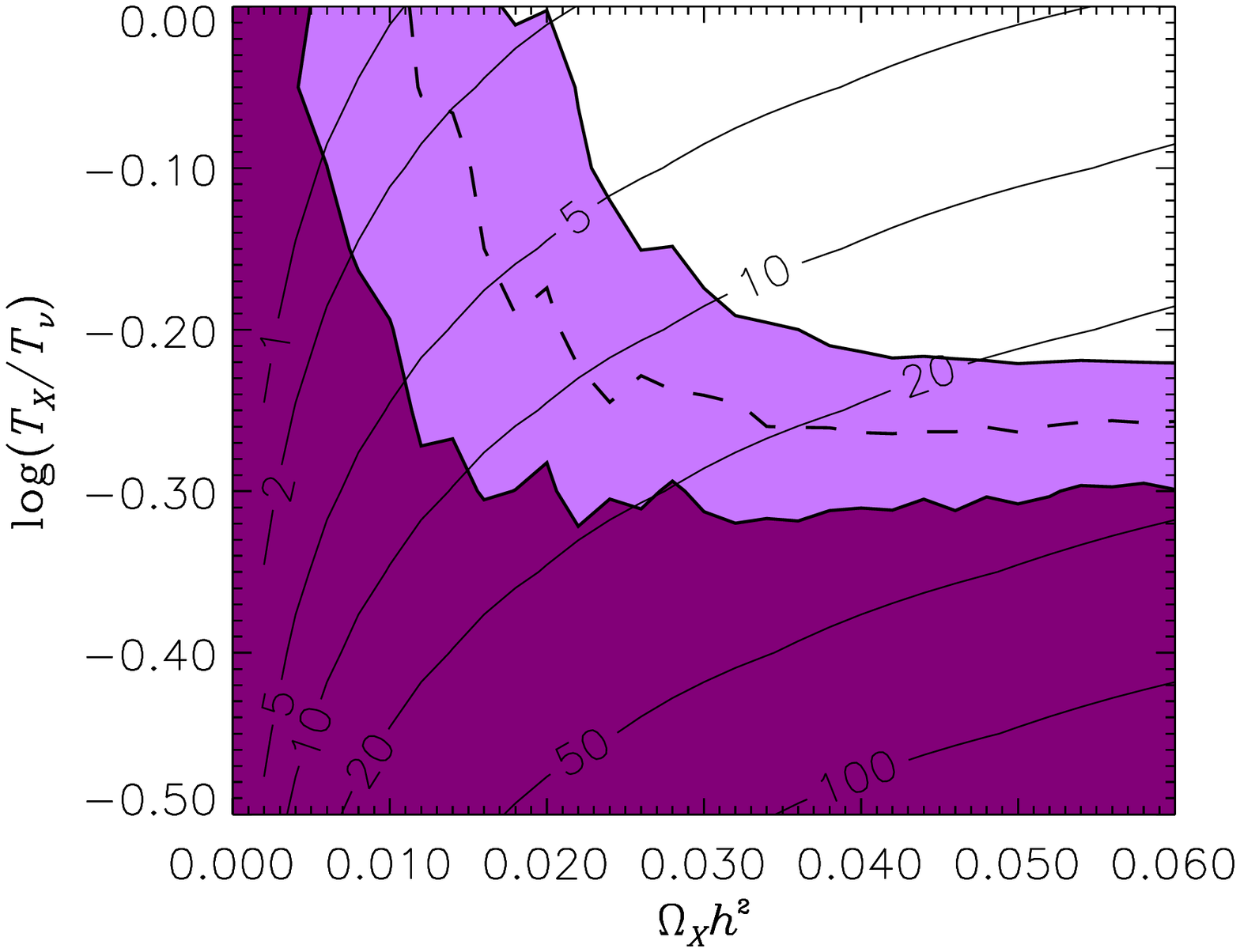}%
\includegraphics[width=70mm]{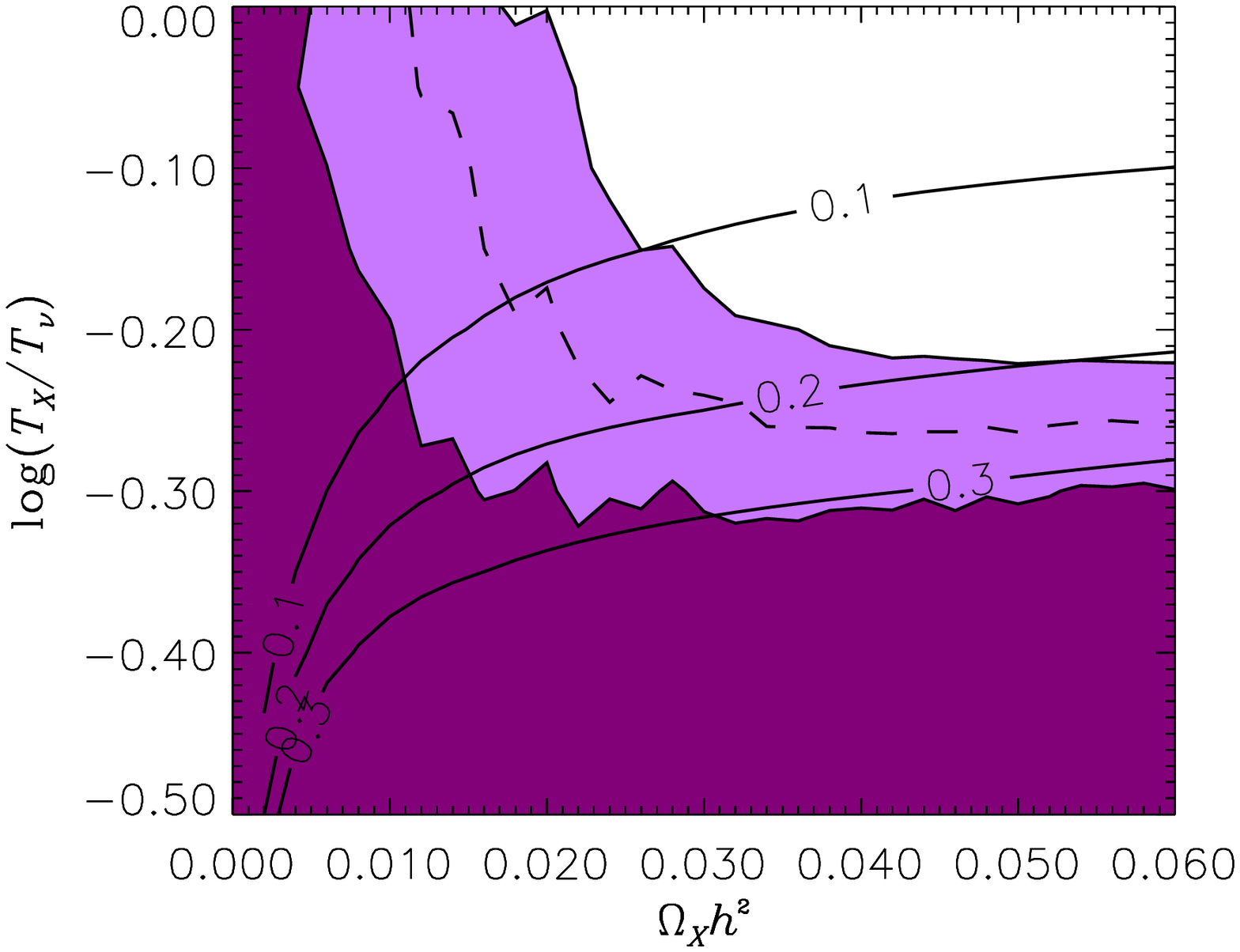}
\caption{Likelihood contours (68\% and 95\%) for a generic thermal
relic with $g_X=2$, characterized by its density, $\Omega_X h^2$, and
effective temperature, $T_X$.  The dashed line shows $\Delta \chi^2 =
4$, corresponding to the 95\% confidence limit for {\it fixed} $T_X$.
In the left panel, the solid lines correspond to the indicated values
of the particle mass in eV, whereas in the right panel they denote the
parameters for which the free streaming scale according to
Eq.~(\ref{eq:freestream}) is $k_{\rm FS}=0.1$, 0.2, and
$0.3\,$\hMpc.}
\label{fig:generic1_1}
\end{center}
\end{figure}

Our first case is $g_X=2$, i.e.\ a Majorana fermion with interactions
that can be weaker than those of ordinary neutrinos. The limits are
calculated with a Fermi-Dirac distribution, but they would not change
significantly for the Bose-Einstein case.  In
Fig.~\ref{fig:generic1_1} we show the 68\% and 95\% likelihood
contours for the allowed parameters in terms of $\Omega_X h^2$ and
$T_X/T_\nu$. Limits for selected values of $T_X/T_\nu$ can be found in
Table~\ref{table:mass}.  Of course, in the limit $T_X/T_\nu=1$ we
recover the 3+1 scenario of 3 ordinary massless neutrinos and one
extra flavor with mass $m_\nu$.

Figure~\ref{fig:generic1_1} reveals that the bound on $\Omega_X h^2$
is almost independent of $T_X$ down to a critical value.  For lower
$T_X$ the bound abruptly relaxes, allowing for a large mass fraction
$\Omega_X$ of the new particles.  This behavior is explained by our
non-linearity cut in the data at $k=0.2\,$\hMpc. For low values of
$T_X$ the free-streaming scale of the particles is simply too small to
affect large-scale structures above the cut.

In order to substantiate this interpretation we note that the
free-streaming scale can be approximated as
\begin{equation}
\lambda_{\rm FS}=\frac{2\pi}{k_{\rm FS}}
\sim \frac{20~{\rm Mpc}}{\Omega_X h^2}
\left(\frac{T_X}{T_\nu}\right)^4 \left\{1+\log \left[3.9\,
\frac{\Omega_X}{\Omega_m} \left(\frac{T_\nu}{T_X}\right)^2\,
\right]\right\}\,. \label{eq:freestream}
\end{equation}
In the right panel of Fig.~\ref{fig:generic1_1} we show contours for
which $k_{\rm FS}= 0.1$, 0.2, and $0.3\,$\hMpc.  Evidently $k_{\rm
FS}=0.2\,$\hMpc\ corresponds fairly well to the value of $T_X$ where
the constraint on $\Omega_X h^2$ disappears.  An exact match can not
be expected because our estimate was only approximate.  Moreover, the
$X$-particles follow a thermal distribution so that some fraction will
have large energies and therefore stream further than the mean of the
distribution.  Therefore, there is some free-streaming effect above
our cut even if the mean free streaming scale is too small.

\begin{table}
\caption{Limits (95\% CL) for
$g_X=2$, i.e.~a general Majorana fermion.}
\begin{indented}
\item[]
\begin{tabular}{@{}lllll}
\br
&&&\multicolumn{2}{l}{Upper limit}\\
$\log(T_X/T_\nu)$ &$T_X/T_\nu$ &$g_{*\rm D}/g_{*\nu}$&
$\Omega_Xh^2$&$m_X$ [eV]\\
\mr \kern0.7em$0.00$&1.000&1.00&0.011& 1.0 \cr
$-0.05$ &0.891&1.41&0.012& 1.6 \\
$-0.10$ &0.794&2.00&0.015& 3.0 \\
$-0.15$ &0.708&2.82&0.017& 4.1 \\
$-0.20$ &0.631&3.98&0.021& 7.5 \\
$-0.25$ &0.562&5.62&0.029& 15.0\\
$-0.30$ &0.501&7.94&\multicolumn{2}{l}{No bounds.}\\
\hline
\end{tabular}
\end{indented}
\label{table:mass}
\end{table}

We note that even though the galaxy data from surveys such as 2dFGRS
or SDSS must be cut at $0.2\,$\hMpc, data from the Lyman-$\alpha$
forest measure an earlier cosmological epoch and thus the linear
regime extends to much larger values of $k$. For instance, this was
used in Ref.~\cite{Narayanan:2000tp} to constrain the possible mass of
a thermally produced warm dark matter particle to be $m_X \gtrsim
750$~eV, assuming that it provides all the dark matter.  In our paper
we have deliberately avoided the Lyman-$\alpha$ data since the
conversion of the measured flux power spectrum into a matter power
spectrum is fraught with difficulties and the result is at present
highly controversial.

For a particle freezing out after the QCD phase transition, the
effective number of thermal degrees of freedom at decoupling is
$g_{*D}\lesssim20$ or $g_{*D}/g_{*\nu}\lesssim 2$ so that the mass
bounds are quite restrictive, $m_X<3$~eV or better. However, at the
QCD phase transition, $g_*$ decreases dramatically due to the large
number of confined colored degrees of freedom. At epochs shortly
before the transition $g_*\gtrsim60$ so that a particle freezing out
at that time is characterized by $g_{*D}/g_{*\nu}\gtrsim 6$,
preventing us from stating any significant bounds without using the
Lyman-$\alpha$ data.

\subsection{Scalar Bosons}

The cosmic mass fraction of our new particles is proportional to
$g_Xm_X$ so that the limits of the previous section can be roughly
scaled to cases with a different number of internal degrees of freedom
$g_X$. Also, on the basis of the estimated free-streaming scale one
can estimate for which value of $T_X/T_\nu$ the mass limits begin to
fail.  Still, we have performed an explicit analysis for another
special case, the one of a scalar or pseudoscalar boson
($g_X=1$). This example is of particular interest for axions or
axion-like particles.

\begin{figure}[t]
\begin{center}
\hspace*{1.5cm}
\includegraphics[width=70mm]{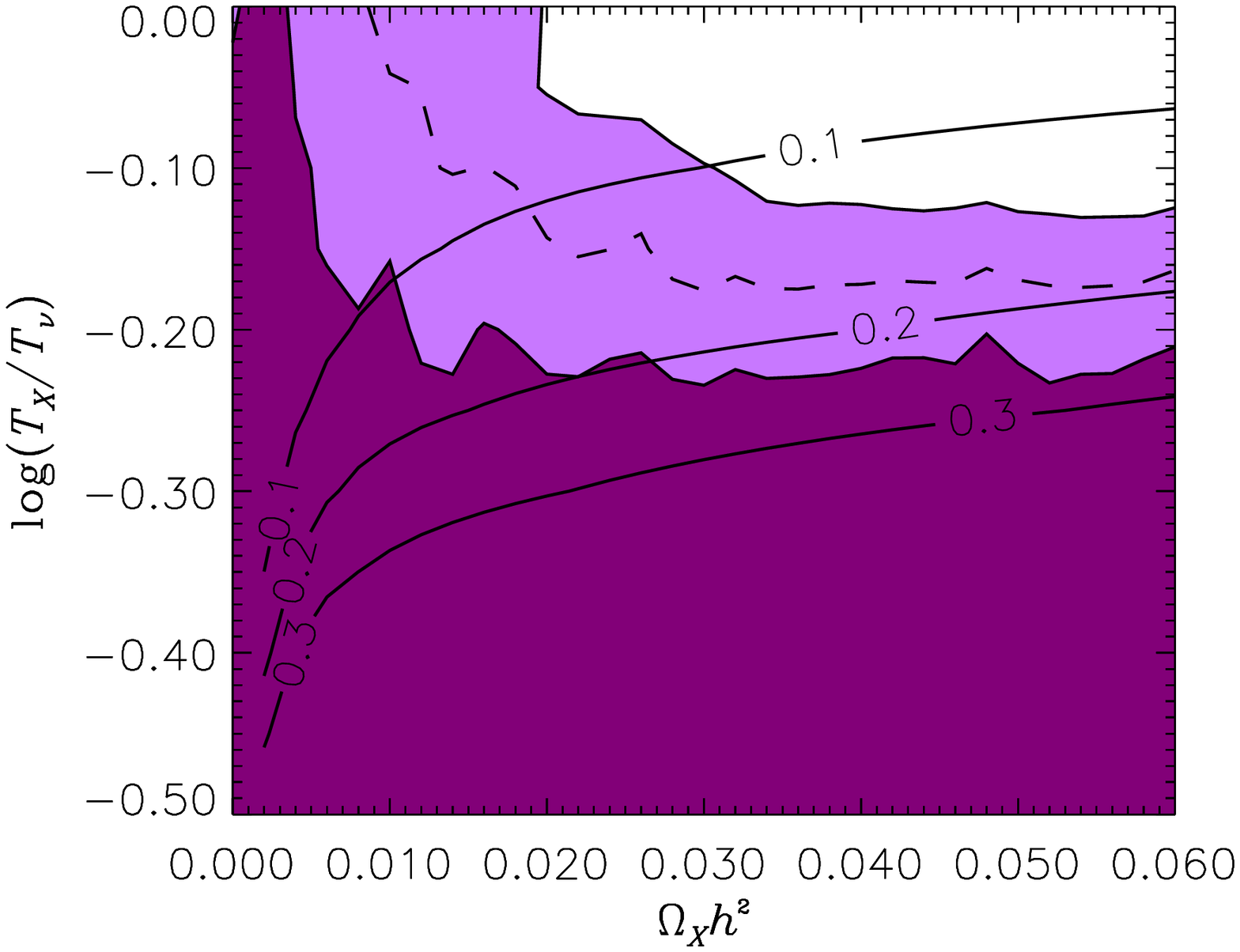}%
\includegraphics[width=70mm]{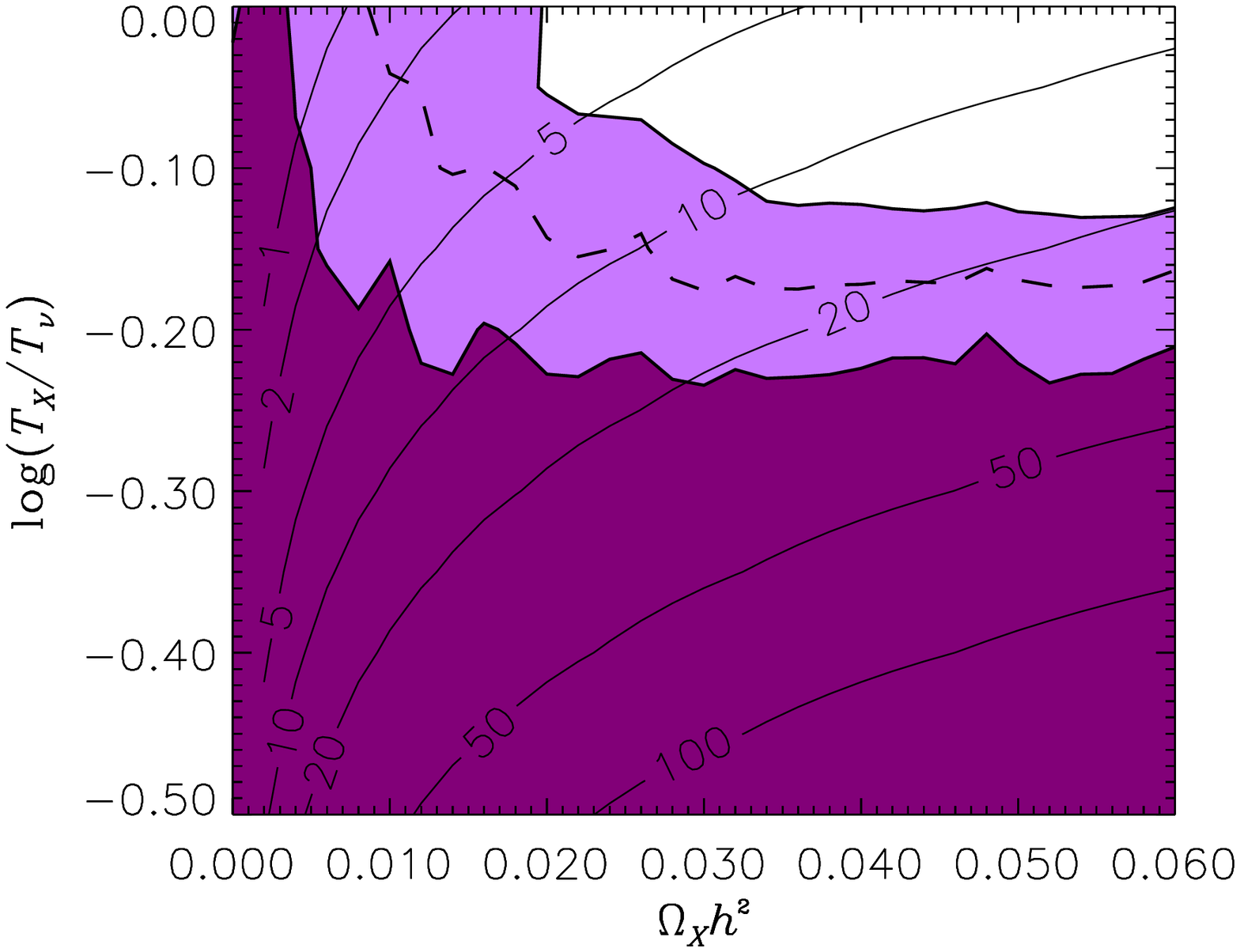}
\caption{Same as Fig.~\ref{fig:generic1_1} for scalar bosons
($g_X=1$).\label{fig:generic1_2}}
\end{center}
\end{figure}

The results of our likelihood analyis are shown in
Fig.~\ref{fig:generic1_2}, which is fully analogous to
Fig.~\ref{fig:generic1_1}. We also tabulate explicit $m_X$ limits for
selected values of $T_X/T_\nu$ in Table~\ref{table:mass2}.  We note
that for particles decoupling after the QCD phase transition the mass
limit for equivalent decoupling epochs is only slightly weaker than it
is for the previous $g_X=2$ case, i.e.\ the simple $m_X\propto
g_X^{-1}$ scaling would have yielded less restrictive limits.  The
free-streaming scale enters the non-linear regime at higher $T_X$ than
before.  Once more we find that there is no meaningful mass limit for
particles decoupling before the QCD~epoch.

\begin{table}
\caption{Limits (95\% CL) for
$g_X=1$, i.e.~a general scalar or pseudoscalar boson.}
\begin{indented}
\item[]
\begin{tabular}{@{}lllll}
\br
&&&\multicolumn{2}{l}{Upper limit}\\
$\log(T_X/T_\nu)$ &$T_X/T_\nu$ &$g_{*\rm D}/g_{*\nu}$&
$\Omega_Xh^2$&$m_X$ [eV]\\
\mr
\kern0.7em$0.00$&1.000&1.00&0.009& 1.2 \\
$-0.05$ &0.891&1.41&0.011& 2.0 \\
$-0.10$ &0.794&2.00&0.013& 3.3 \\
$-0.15$ &0.708&2.82&0.020& 7.8 \\
$-0.20$ &0.631&3.98&\multicolumn{2}{l}{No bounds.}\\
\hline
\end{tabular}
\end{indented}
\label{table:mass2}
\end{table}

As a concrete example we mention axions. The energy-loss limit from
SN~1987A implies that the axion mass is bounded by $m_a\lesssim
10^{-2}$~eV \cite{Raffelt:1999tx}. Therefore, it is usually assumed
that cosmologically significant axions must be non-thermal relics with
very small masses. On the other hand, the SN~1987A argument may have
loop holes or one may assume that the axion couplings, notably those
to photons, are quite different from the most common models, perhaps
allowing eV-mass axions to exist in the ``hadronic axion window''
\cite{Moroi:1998qs}. If the axion mass is in the eV range, its
couplings to nucleons and pions is so strong that it decouples
thermally well after the QCD
transition~\cite{Turner:1986tb,Chang:1993gm,Masso:2002np}.  According
to Table~\ref{table:mass2} this implies a cosmological mass limit of
$m_a<2$--3~eV.

In a direct telescope search, narrow lines from the two-photon decay
of axions trapped in galaxy clusters were searched in the mass range
3--8~eV \cite{Bershady:1990sw}, but no signal was found.  According to
our new limits, axions in this mass range can not exist because of
their excessive free-streaming effect on cosmic structures.

On the other hand, the possibility of axion-like particles that couple
only to photons by virtue of their $\pi^0$-like two-photon vertex is
sometimes discussed~\cite{Masso:1995tw,Masso:1997ru}.  For an
interaction strength small enough to obey the well-known
globular-cluster limit~\cite{Raffelt:1999tx}, such particles would
decouple well before the QCD epoch~\cite{Masso:1995tw}. Therefore, our
arguments do not provide any new limits on such hypothetical
particles.

Either way, our limits do not infringe on the possibility that the
CAST experiment at CERN to search for solar axions could discover
axions or axion-like particles with masses up to the eV range
\cite{Collar:2003ik}. Actually, a small gap will remain between our
new limits and CAST's future high-mass frontier.  Likewise, the future
possibility to search for axion-like particles with a string of
decommissioned HERA magnets~\cite{Ringwald:2003ns} remains~unaffected.

\section{Discussion and Summary}

We have calculated general cosmological bounds on the masses and
abundances of light, stable particles such as neutrinos and
axions. For neutrinos we have calculated three different generic
cases:

{1) All neutrinos are massive, so that there are $N_\nu$
neutrinos, each with mass $m_\nu$.}

{2) Only one neutrino is massive, but altogether there
are $N_\nu$ neutrinos.}

{3) There are three massless neutrinos and $N_\nu-3$
species with mass $m_\nu$.}

\noindent In each case it is found that there is a strong correlation
between $\Omega_\nu h^2$ and $N_\nu$, but not necessarily between
$m_\nu$ and $N_\nu$.

We have extended the standard analysis to include generic light
particles such as axions and majorons. We assume such particles to
have been thermally produced so that they are characterized by an
effective temperature $T_X$. The other free parameter can be taken to
be either $\Omega_X h^2$ or $m_X$, but since $\Omega_X h^2$ is the
main parameter entering into CMB and LSS calculations we chose the
parameters to be $\Omega_X h^2$ and $T_X$.

In our analysis we do not use LSS data in the non-linear regime, i.e.\ 
at small scales, and we also do not use Lyman-$\alpha$ data.
Therefore, particles that decouple too early and thus today have much
smaller momenta than neutrinos do not produce significant
free-streaming effects on the relevant scales. For both
spin-$\frac{1}{2}$ fermions and for scalars that decouple before the
QCD epoch, we can not state a significant mass limit.  On the other
hand, for such particles decoupling after the QCD epoch, there is a
mass limit between 1--3~eV, depending on the spin and exact epoch of
decoupling. For axions, our limits supersede previous results from
direct telescope searches for axion decays.

An independent study of cosmological bounds on neutrino masses and
relativistic relics was completed shortly after
ours~\cite{Crotty:2004gm}. For overlapping cases there is agreement
with our work.

\section*{Acknowledgments} 

We acknowledge use of the publicly available CMBFAST
package~\cite{CMBFAST} and of computing resources at DCSC (Danish
Center for Scientific Computing).  We acknowledge support by the
European Science Foundation (ESF) under the Network Grant No.~86
Neutrino Astrophysics. In Munich, this work was supported, in part, by
the Deutsche Forschungsgemeinschaft (DFG) under grant No.~SFB-375.

\vspace*{2cm}

\section*{References} 

\end{document}